\documentclass[onecolumn,superscriptaddress,floatfix,showpacs]{revtex4-2}

\usepackage{tabularx}

\usepackage[utf8]{inputenc}
\usepackage[T1]{fontenc}     
\usepackage[british]{babel}  
\usepackage[sc,osf]{mathpazo}
\usepackage[scaled=0.86]{berasans}  
\usepackage[colorlinks=true, citecolor=blue, urlcolor=blue]{hyperref}  
\usepackage{mathtools}
\usepackage{lipsum}
\usepackage{graphicx} 
\usepackage{subfig}
\usepackage[babel]{microtype}  
\usepackage{amsmath,amssymb,amsthm,bm,amsfonts,mathrsfs,bbm} 

\usepackage{xspace}  
\usepackage{pgfplots}
\usepackage{xcolor,colortbl}
\def\ba{\begin{equation}}
\def\ea{\end{equation}}
\def\bea{\begin{eqnarray}}
\def\eea{\end{eqnarray}}
\def\ben{\begin{equation*}}
\def\een{\end{equation*}}
\def\bean{\begin{eqnarray*}}
\def\eean{\end{eqnarray*}}
\def\bma{\begin{mathletters}}
\def\ema{\end{mathletters}}
\def\bi{\begin{itemize}}
\def\ei{\end{itemize}}

\newcommand{\be}{\begin{equation}}
\newcommand{\ee}{\end{equation}}

\newcommand{\kommentar}[1]{}

\newcommand{\forget}[1]{}

\newtheorem{theorem}{Theorem}

\begin{document}

\title{Quantum conditional entropies and steerability of states with maximally mixed marginals}

\author{Komal Kumar}
\email{p20210063@hyderabad.bits-pilani.ac.in}
\affiliation{Department of Mathematics, Birla Institute of Technology and Science Pilani, Hyderabad Campus,Telangana-500078, India}
\author{Nirman Ganguly}
\email{nirmanganguly@hyderabad.bits-pilani.ac.in}
\affiliation{Department of Mathematics, Birla Institute of Technology and Science Pilani, Hyderabad Campus,Telangana-500078, India}

\begin{abstract}
Quantum steering is an asymmetric correlation which occupies a place between entanglement and Bell nonlocality. In the paradigmatic scenario involving the protagonists Alice and Bob, the entangled state shared between them, is said to be steerable from Alice to Bob if the steering assemblage on Bob's side do not admit a local hidden state (LHS) description. Quantum conditional entropies, on the other hand provide for another characterization of quantum correlations. Contrary to our common intuition conditional entropies for some entangled states can be negative, marking a significant departure from the classical realm.  Quantum steering and quantum nonlocality in general share an intricate relation with quantum conditional entropies. In the present contribution, we investigate this relationship. For a significant class, namely the two-qubit Weyl states we show that negativity of conditional Rényi 2-entropy and conditional Tsallis 2-entropy is a necessary and sufficient condition for the violation of a suitably chosen three settings steering inequality. With respect to the same inequality we find an upper bound for the conditional Rényi 2-entropy, such that the general two-qubit state is steerable.  Moving from a particular steering inequality to local hidden state descriptions, we show that some two-qubit Weyl states which admit a LHS model possess non-negative conditional Rényi 2-entropy. However, the same does not hold true for some non-Weyl states. Our study further investigates the relation between non-negativity of conditional entropy and LHS models in two-qudits for the isotropic and Werner states. There we find that whenever these states admit a LHS model, they possess a non-negative conditional Rényi 2-entropy. We then observe that the same holds true for a noisy variant of the two-qudit Werner state.   
\end{abstract}
\date{\today}
\maketitle

\section{Introduction}
The study on correlations unachievable within the classical paradigm has been an integral constituent of theoretical \cite{P.Zoller} and experimental \cite{P.Zoller} probes in quantum information processing. Entanglement \cite{R1.Horodecki}, is the supreme contributor to such studies. However, entanglement although a form of quantum nonlocality, is not the strongest among such correlations. Bell nonlocality \cite{J.Bell} constitutes a stronger correlation significant both from a foundational \cite{V.Scarani} and pragmatic perspective \cite{V.Scarani}. The recent spurt in device-independent security protocols \cite{A.Acin} has provided for a driving force to studies on Bell nonlocality. 

Quantum steering \cite{R.Uola} on the other hand lies between these two form of correlations and is usually asymmetric in nature. Schrödinger was the first to notice this phenomenon \cite{E.Schrodinger1,E.Schrodinger2} and considered it to be intriguing that the first party can by choosing a measurement steer the state on the other side into an eigenstate of position or momentum. However, apart from being a thought-provoking fundamental question this phenomenon didn't receive much attention till the work of Wiseman, Jones and Doherty \cite{H.Wiseman}. Although Schrödinger first employed the term "steering" in this context, it did not refer to the modern and precisely defined notion of quantum steering. The authors in \cite{H.Wiseman}  formulated steering in the language of quantum information processing and gave an operational interpretation to it \cite{H.Wiseman}. In the modern perspective steering denotes the impossibility to describe the conditional states at one party by a local hidden state (LHS) model. Much like Bell nonlocality, quantum steering also has gained prominence recently due to its significance in semi-device-independent communication scenarios \cite{M.Pawlowski}. An interesting observation here is that although entanglement is necessary for steering and Bell nonlocality, it is not sufficient.

In any theory of information processing, entropies \cite{A.Wehrl} play a key role as they prove to be efficient quantifiers of information content. The Shannon entropy \cite{C.E.Shannon} in classical information theory and its quantum counterpart, the von Neumann entropy \cite{J.Neumann} are the paradigmatic examples. However, these are not the only versions of entropy to be used. Rényi $\alpha$- entropies \cite{AR}, Tsallis $\alpha$- entropies \cite{TS} and their quantum analogues \cite{M.Muller} are  used in different scenarios. In quantum information theory, we broadly distinguish tasks as falling into two regimes:(i) asymptotic regime, where many identically and independently distributed quantum systems appear and (ii) single shot regime, where scenarios involve only a single quantum system. While in (i) von Neumann entropies play a significant role, in (ii) it is the Rényi $\alpha$- entropy that takes center stage  \cite{M.Muller}. Quantum entropies have not only played their part in information processing, but also have been used in the establishment of entropic Bell inequalities \cite{N.Cerf}, entropic steering inequalities \cite{J.Schneeloch,AC} and entropic uncertainty relations \cite{S.Wehner} which in turn is at the heart of quantum cryptography \cite{N.Gisin}. 

Quantum entropies generalizes classical entropies in a way and thus it is a function of the density matrix. In this perspective, consider a bipartite system shared between Alice and Bob and ask the question "\textit{What amount of information Bob has to learn in addition to the knowledge he already possesses to know the whole system ?}". The answer to this question is provided by conditional entropies. In the classical realm, conditional entropies are always nonnegative in consonance to our intuition. However, in the quantum situation conditional entropies can be negative \cite{N.J.Cerf} marking a radical departure from classical notions. In fact certain entangled states possess negative conditional entropy. An operational interpretation to quantum conditional von Neumann entropy was given by the authors in \cite{M.Horodecki} akin to the classical Slepian-Wolf theorem \cite{SLP}. Recently the negativity of conditional von Neumann entropy was identified as a resource \cite{M1.Vempati,M2.Vempati,MW}. In another recent work \cite{Gourinev}, it was noted that in any axiomatic characterization of conditional entropy, this negativity is inevitable. 

The present work investigates the relation between these two extremely significant measures of quantum correlations, namely steering and conditional entropies. Probes were made to see this in case of Bell-CHSH inequalities \cite{J.F.Clauser} and conditional von Neumann entropies in two-qubit systems \cite{N.Friis}. The relation of Bell's inequalities with quantum $ \alpha $ entropies were investigated in \cite{R.Horodecki}. As noted before, entropies were also used to form entropic Bell's inequalities \cite{N.Cerf} and entropic steering inequalities \cite{J.Schneeloch}. However, to the best of our knowledge, these studies have not been done in the framework of steering and LHS models. Our work tries to bridge this gap. 

We start with two-qubit systems and probe the relation of conditional entropies with the violation of a three-settings steering inequality, quite prominently known as the Cavalcanti, Jones, Wiseman , Reid (CJWR) inequality \cite{E.Cavalcanti}. We show that for the Weyl states(states with maximally mixed marginals), the negativity of conditional Rényi 2-entropy is necessary and sufficient for the violation of the CJWR inequality. Upper bounds to the conditional entropy of generic two-qubit quantum states are also provided which guarantees the violation of the inequality. We then move from a particular steering inequality to LHS models. We use a sufficient criteria for unsteerability \cite{J.Bowles} and find that whenever Weyl states are unsteerable due to the criteria , they possess a non-negative conditional Rényi 2-entropy. However, some non-Weyl states present a different picture. We extend our study to two-qudit systems and observe the behavior of LHS models of the isotropic and Werner states \cite{R.Werner} vis a vis conditional Rényi 2-entropies. We observe, whenever the states admit a LHS model, they have non-negative conditional Rényi 2-entropy. We then introduce white noise in the Werner states and note that the LHS model of the transformed state exhibits the same characteristic.

The paper is arranged as follows: In the next section we revisit some key notions important for our work and fix the notations. In section(\ref{CJWRtwo}), we study the relation between violation of the CJWR inequality and conditional entropy of two-qubit systems. In section(\ref{LHSqubit}), we proceed to investigate the LHS models in two-qubit systems. We then explore the relation with LHS models in two-qudits in section(\ref{LHSqudit}). We then end with the concluding remarks.

\section{Notation and Preliminaries}
In this section we fix the notations and recapitulate some notions important for the study. \\
Our work is done in finite dimensional Hilbert spaces. Quantum states are described by density operators $ \rho $, i.e., positive semi-definite ($ \rho \ge 0 $), Hermitian (which also follows from positivity) with unit trace. $ \rho \in \mathfrak{B}(\mathbb{H}_{A} \otimes \mathbb{H}_{B}) $, where $ \mathfrak{B}(\mathbb{H}_{A} \otimes \mathbb{H}_{B}) $ (also known as the Hilbert-Schmidt space) represents the bounded linear operators acting on $ \mathbb{H}_{A} \otimes \mathbb{H}_{B} $. $ \mathbb{H}_{X},X=A,B $ denote the underlying Hilbert space. $ S(.) , S_\alpha(.), S^T_\alpha(.) $ denote respectively the von Neumann, Rényi $ \alpha- $ entropy and Tsallis $ \alpha- $ entropy of the quantum state being discussed.
\subsection{Bloch-Fano Decomposition of density matrices}

For bipartite quantum systems, the density matrices can be represented as \cite{N.Friis}
\begin{widetext}
\begin{equation} \label{blochqudit}
	\rho_{d_A \otimes d_B} = \frac{1}{d_A d_B} [\mathbb{I}_A \otimes \mathbb{I}_B + \sum_{m=1}^{d_A^2 - 1} a_m g_m^A \otimes \mathbb{I}_B + \sum_{n=1}^{d_B^2 - 1} b_n  \mathbb{I}_A \otimes g_n^B + \sum_{m=1}^{d_A^2 - 1} \sum_{n=1}^{d_B^2 - 1} t_{mn} g_m^A \otimes g_n^B],
\end{equation}
\end{widetext}
 where $ \text{dim}~ \mathbb{H}_{A} = d_A  $ and $ \text{dim}~ \mathbb{H}_{B} = d_B  $. The hermitian operators $ g_m^i $ for $ i=A,B $ are generalizations of the Pauli matrices, i.e., they are orthogonal $ Tr[g_m^i g_n^i] = 2 \delta_{mn} $ and traceless, $ Tr[g_m^i] =0  $ and for single qubit systems they coincide with the Pauli matrices. The coefficients $ a_m, b_n \in \mathbb{R} $ are the components of the generalized Bloch vectors $ \vec{a} , \vec{b} $ of the subsystems $ A,B $ respectively. The real coefficients $ t_{mn} $ are the components of the correlation tensor. For two-qubit systems the density matrices can be represented as, 
 \begin{equation}\label{blochqubit}
 	\rho_{2 \otimes 2}= \frac{1}{4} [\mathbb{I}_{2}\times\mathbb{I}_2+\vec{a}.\vec{\sigma}\otimes \mathbb{I}_2+\mathbb{I}_2\otimes \vec{b}.\vec{\sigma} + \sum_{m=1}^{3} \sum_{n=1}^{3} t_{mn} \sigma_m^A \otimes \sigma_n^B ].
 \end{equation} 
 Here $ \vec{\sigma} $ refers to the vector of qubit Pauli matrices. \\
 An interesting class of states is the \textit{locally maximally mixed states or Weyl states} which in the two-qudit systems (up to local unitaries) is given by, 
 
 \begin{equation}
 	\rho^{weyl}_{d} = \frac{1}{d^2} [\mathbb{I}_A \otimes \mathbb{I}_B + \sum_{n=1}^{d^2 - 1} w_{n} g_n^A \otimes g_n^B].
 \end{equation}
 The reduced marginals of such states are maximally mixed, i.e., $ Tr_A[\rho^{weyl}_{d}] = \mathbb{I}_A/d  $ and $ Tr_B[\rho^{weyl}_{d}] = \mathbb{I}_B/d.$

\subsection{Quantum Entropies}
The von Neumann entropy of a quantum state $ \rho $ is defined as,
\begin{equation}
	S(\rho) = -Tr [ \rho \log \rho].
\end{equation}
where the logarithms are taken to the base 2.
The von Neumann entropy has a special association with the eigenvalues of the density matrix, i.e., it is a function of the eigenvalues. The corresponding conditional von Neumann entropy(CVNE) is given by $ S(A|B)=S(\rho)-S(\rho_B) $.

The Rényi $\alpha$-entropy is given by
\begin{align}
	S_{\alpha}(\rho)= \frac{1}{1-\alpha} \log[Tr(\rho^{\alpha})], \alpha >0, \alpha \neq 1.
\end{align}
The von Neumann entropy is the limiting case of the Rényi $\alpha$-entropy as 
$\alpha\rightarrow 1$. The corresponding conditional Rényi $\alpha$- entropy (CRAE) is given by $ S_\alpha(A|B)=S_\alpha(\rho)-S_\alpha(\rho_B) $.

Tsallis $\alpha$-entropy is given by 
\begin{align}
	S^T_{\alpha}(\rho)= \frac{1}{1-\alpha} [Tr(\rho^{\alpha})-1], \alpha >0, \alpha \neq 1.
\end{align}
The corresponding conditional Tsallis $\alpha$-entropy (CTAE) is given by \cite{Vollbrecht}, 
\begin{align*}
	S^T_{\alpha}(A|B)=\frac{Tr(\rho_B^\alpha)-Tr(\rho^\alpha)}{(\alpha-1)Tr(\rho_B^\alpha)}.
\end{align*}

\subsection{Quantum Nonlocality}
\textit{Bell nonlocality-} Consider two parties Alice and Bob to be spatially separated, sharing some quantum state $\rho_{AB}$. They perform measurements denoted by $ A_x , B_y $ respectively on their particles, the inputs being labelled by $ x $ and $ y $. The corresponding outcomes at Alice's and Bob's side are labelled as $ a $ and $ b$. Now, for some hidden variable $\lambda$ \& probability distribution $q(\lambda)$, if all joint probabilities $p(a,b|x,y)$ can be expressed as, 
\begin{align} \label{LHV}
	p(a,b|x,y) = \int_{\lambda} d\lambda q(\lambda)p(a|x,\lambda)p(b|y,\lambda)
\end{align}
then we say that the state is local or has a local hidden variable (LHV) model.  
Here, $p(a|x,\lambda)$ \& $p(b|y,\lambda)$ are said to be the local response functions of Alice and Bob respectively and $ q(\lambda) $ is the probability distribution of the hidden variable $ \lambda $. If equation (\ref{LHV}) does not hold, we say that it is nonlocal and hence will violate a suitably chosen Bell's inequality \cite{J.Bell}. \\
\textit{Bell-CHSH Criterion}- The Bell-CHSH criterion \cite{J.F.Clauser} gives a Bell's inequality in the two-qubit scenario when there are two parties performing  measurements in a two input-two output scenario. A necessary and sufficient condition for its violation was provided in \cite{HoroBell}. Let $\rho$ be a two-qubit density operator and its correlation tensor be given by $T = [t_{mn}]$. Suppose that the largest two eigenvalues of matrix $T^{\dagger}T$ are $\lambda_{1}$ \& $\lambda_{2}$. Then the state $\rho$ violates the CHSH inequality iff their sum is greater than one, 
i.e., $\lambda_{1}$+$\lambda_{2}$ $> 1$ \cite{HoroBell}. One should note here that a state may satisfy the Bell-CHSH criterion, however it may still be nonlocal (through the violation of any other Bell's inequality).\\
\textit{Quantum Steering-} Quantum steering is an asymmetric form of nonlocality, counted on the ability of Alice to steer the state on Bob's side with a choice of measurement on her side. Although the concept of steering is not new, dating back to the contributions of Schrödinger \cite{E.Schrodinger1,E.Schrodinger2}, the operational formulation was provided only in \cite{H.Wiseman}. The notion of steering can be conveniently expressed in the language of steering assemblages. Suppose that there are two parties Alice and Bob spatially separated, sharing quantum state $\rho_{AB}$. Now, Alice performs some type of measurement, say $x$ on her part of the state and the outcome obtained is $a$. Bob remains with unnormalized conditional state $\rho_{a|x}$ (steering assemblage) for each measurement of Alice. Now if Bob's conditional state $\rho_{a|x}$ can be expressed as,
\begin{align}
	\rho_{a|x}= \int_{\lambda} d\lambda p(\lambda)p(a|x,\lambda)\sigma_{\lambda},
\end{align}
where $\sigma_{\lambda}$ is the hidden state (on Bob's side) with probability $p(\lambda)$ and $p(a|x,\lambda)$ is the local response function of Alice,
then we say that the state $\rho_{AB}$ is unsteerable or has local hidden state (LHS) model. Otherwise, the state $\rho_{AB}$ is steerable from Alice to Bob. Steerability is an asymmetric trait. A state which is steerable from Alice to Bob might not be steerable from Bob to Alice.\\
\textit{The CJWR inequality for steering-} Cavalcanti, Jones, Wiseman and Reid(CJWR) derived a
series of correlators based inequalities \cite{E.Cavalcanti} for verifying steerability of $\rho:$
\begin{equation}\label{mon2}
	F_{n}(\rho,\nu) = \frac{1}{\sqrt{n}}|\sum_{l=1}^{n} \langle A_{l} \otimes B_{l} \rangle | \leq 1,
\end{equation}
with,
\begin{equation}
A_{l} = \hat{u}_{l} \cdot \overrightarrow{\sigma},B_{l}=\hat{v}_{l} \cdot \overrightarrow{\sigma}
\end{equation}
 where $\hat{u}_{l} $$\in$$ \mathbb{R}^{3} \textmd{are unit vectors whereas}\,\hat{v}_{l} $$\in$$ \mathbb{R}^{3}\,$denote orthonormal vectors$.\,$$\nu $$=$$\{\hat{u}_{1},\hat{u}_{2},....\hat{u}_{n}, \hat{v}_{1},\hat{v}_{2},...,\hat{v}_{n} \}$ stands for the collection of measurement directions,  $\langle A_{l} \otimes B_{l} \rangle = \textmd{Tr}(\rho (A_{l} \otimes B_{l}))$. Violation of Eq.(\ref{mon2}) ensures both way steerability of $\rho$ in the sense that it is steerable from $A$ to $B$ and vice versa. In particular, for $n$$=$$3,$ CJWR inequality(Eq.(\ref{mon2})) for three settings takes the form:
\begin{equation}\label{mon2i}
	F_{3}(\rho,\nu) = \frac{1}{\sqrt{3}}|\sum_{l=1}^{3} \langle A_{l} \otimes B_{l} \rangle | \leq 1,
\end{equation}

In \cite{Costa}, analytical expression for the violation of the linear inequality (equation(\ref{mon2i})) was given. Any two-qubit state shared between $A$ and $B$ is both-way $F_{3}$ steerable if:
\begin{align}
	\sqrt{Tr(T^{\dagger}T)}> 1,
\end{align}
$ T $ being the correlation tensor.

 One must note here that entanglement is only a necessary criterion for Bell nonlocality and steering.  
\section{Conditional Entropies and Steerability for two-qubit states} \label{CJWRtwo}
In this section we investigate the relation between conditional entropies and $ F_3 $ steerability for two-qubit Weyl states. We start our discussion with the Werner state, which is a special kind of Weyl state. 
\subsection{Werner States}

\textit{Relation with conditional von Neumann entropy-} Consider the two-qubit Werner state $\rho^{wer}_{2}(p)=$ $\frac{1}{4}[\mathbb{I}_2 \otimes \mathbb{I}_2-\sum_{i=1}^{3}p(\sigma_{i} \otimes \sigma_{i})]$. Conditional von Neumann entropy is given by, 
\begin{widetext}
\begin{align}
	S(A|B)_{\rho^{wer}_{2}(p)}=-3\left( \frac{1-p}{4}\right) \log\left(\frac{1-p}{4}\right)-\left(\frac{1+3p}{4}\right)\log\left(\frac{1+3p}{4}\right)-1.
\end{align}
\end{widetext}
CVNE is negative for $ p >  0.747614 $. The Werner state is $ F_3 $ steerable for $ p > 0.57735 $. Thus, if the Werner state has negative CVNE then it is 3-steerable.\\

\textit{Relation with other $ \alpha- $ conditional entropies-}
We see here the relation between $ F_3 $ steerability and CRAE for the two-qubit Werner state. The table below notes negativity of CRAE for $ \alpha = 2,3,4,5 $.

\begin{center}
	\begin{tabular}{|c|c|c|c|c|}
	\hline
	$ \alpha$ &  2 &  3 &  4 &  5 \\
	\hline
	Negative CRAE & $\frac{1}{\sqrt{3}}< p$  & $\frac{1}{2}<p$  &  $0.45786<p$ & $0.432041<p$  \\
	\hline
\end{tabular}\\
\end{center}

Note that the Werner state is $ F_3 $ steerable iff the conditional Rényi 2-entropy is negative, a fact that we prove to be true for all two-qubit Weyl states in the next subsection. 
Since  $ S^T_{\alpha}(A|B) \ge 0 \Leftrightarrow S_{\alpha}(A|B) \ge 0  $ \cite{Vollbrecht}, we arrive at the same conclusion for the conditional Tsallis 2-entropy.\\

\subsection{Weyl States}

We consider the Weyl states in two-qubits given by, 
\begin{equation}
\rho^{weyl}_{2}= \frac{1}{4}[\mathbb{I}_2 \otimes \mathbb{I}_2  +\sum_{i=1}^{3} t_{i}(\sigma_{i} \otimes \sigma_{i})].
\end{equation}
The conditional Rényi 2-entropy of $ \rho^{weyl}_{2}  $ is given by $S_2 (A|B)_{\rho^{weyl}_{2}}=1-\log(1+t_{1}^2 + t_{2}^2 + t_{3}^2) $ and $ F_3 (\rho^{weyl}_{2})= \sqrt{t_{1}^2+t_{2}^2+t_{3}^2} $. A simple algebraic calculation shows that $ S_2 (A|B)_{\rho^{weyl}_{2}} < 0 \Leftrightarrow F_3 (\rho^{weyl}_{2}) > 1 $.

We thus have the following theorem, 

\begin{theorem}
	The two-qubit Weyl state is $ F_3 $ steerable iff its conditional Rényi 2-entropy is negative.
\end{theorem}
The same conclusion follows for the conditional Tsallis 2-entropy due to its equivalence with conditional Rényi 2-entropy.\\
\subsection{Two-Qubit general state}
We now consider the general two-qubit state $ \rho^{g}_{2} $, whose decomposition is given in Eq.(\ref{blochqubit}). Its conditional Rényi 2-entropy (CR2E) is given by $S_2 (A|B)_{\rho^{g}_{2}}= \log\left[\frac{2+2||\vec{b}||^{2}}{1+||\vec{a}||^{2}+||\vec{b}||^{2}+||\vec{T}||^{2}}\right]$where $|| . ||$ denote the Euclidean norm, $||T||^{2}=Tr(T^{\dagger} T)$ and it is $F_{3}$ steerable if $||T||=\sqrt{Tr(T^{\dagger}T)}>1$. An algebraic calculation shows that the state is $F_{3}$ steerable $\iff$ $S_2 (A|B)_{\rho^{g}_{2}}< \log\left[\frac{1+||\vec{b}||^{2}}{1+\frac{||\vec{a}||^{2}+||\vec{b}||^{2}}{2}}\right]$.\\
We thus have the following theorem,

\begin{theorem}
	The two-qubit general state is $ F_3 $ steerable iff its CR2E$< \log\left[\frac{1+||\vec{b}||^{2}}{1+\frac{||\vec{a}||^{2}+||\vec{b}||^{2}}{2}}\right]$.
\end{theorem}

\subsection{Absolute non-violation of the CJWR inequality by the Weyl states for a certain purity threshold }
An interesting class of states, which we denote by $ AF_3 $, is the collection of states which preserve $ F_3 $ unsteerability even under global unitary operations. In \cite{S.Bhattacharya}, the authors proved that a quantum state will remain $ F_3 $ unsteerable under arbitrary global unitary operations iff its purity $ \le \frac{1}{2} $. Therefore we have, 
\begin{align}
	Tr[(\rho^{weyl}_{2})^2] \le \frac{1}{2} \nonumber \\
	\Leftrightarrow  1+t_{1}^2 + t_{2}^2 + t_{3}^2 \le 2 \nonumber \\
	\Leftrightarrow S_2 (A|B)_{\rho^{weyl}_{2}} \ge 0.
\end{align} 

In fact the preceding shows that whenever the Weyl states are $ F_3 $ unsteerable, the unsteerability is robust even against global unitary operations.
 However, this observation doesn't hold true for a generic two qubit state, for which we have the following result,

The purity for a general two-qubit state is given by, 
 $Tr[(\rho_{2}^{g})^{2}]=\frac{1}{4}\left[1+||\vec{a}||^{2}+||\vec{b}||^{2}+||\vec{T}||^{2}\right]$ and CR2E is given by $S_{2}(A|B)_{\rho_{2}^{g}}= \log\left[\frac{2+2||\vec{b}||^{2}}{1+||\vec{a}||^{2}+||\vec{b}||^{2}+||\vec{T}||^{2}}\right]$\\
 
 Now, 
$Tr[(\rho_{2}^{g})^{2} \leq \frac{1}{2}$ $\implies$ $S_{2}(A|B)_{\rho_{2}^{g}}\geq 0$. 
We thus have the following theorem
\begin{theorem}
    If $\rho_{2}^{g} \in AF_{3}$ then CR2E is non-negative.
\end{theorem}

\subsection{A non-Weyl state}
We now consider the state $ \rho_2^\theta = \beta | \psi_\theta \rangle \langle \psi_\theta | + (1-\beta) \mathbb{I}_2/2 \otimes \mathbb{I}_2/2$, where $ | \psi_\theta \rangle = \cos \theta |00 \rangle + \sin \theta | 11 \rangle  $ and $ \beta \in [0,1] $. The state is $ F_3 $ steerable and its conditional Rényi 2-entropy is negative when $ \cos4\theta < \frac{2\beta^{2}-1}{\beta^{2}}$ for $\theta \in \left(0,\frac{\pi}{2}\right)\cup\left(\frac{\pi}{2},\pi\right)\cup \left(\pi,\frac{3\pi}{2}\right) \cup \left(\frac{3\pi}{2},2\pi\right)$ and $\beta \in \left(\frac{1}{\sqrt{2-\cos4\theta}},1\right]$. Thus, the state violates the CJWR inequality if and only if it possesses a negative conditional Rényi 2-entropy, for the above restrictions on $ \theta $ and $ \beta $.

\section{Conditional entropies and LHS Models in two-qubits}\label{LHSqubit}

The CJWR inequality being a particular steering inequality, satisfaction of it doesn't imply that the state is unsteerable. There may be other steering inequalities that the state violates. Therefore, it is worthwhile to see the relation between unsteerability (in terms of admitting LHS models) and conditional entropies. 

In \cite{J.Bowles}, a sufficient condition for the unsteerability of states was proposed, which is stated below. A two-qubit state of the form $ \rho_0 = \frac{1}{4} [\mathbb{I}_2 \otimes \mathbb{I}_2 + \vec{\mathfrak{a}}.\vec{\sigma}\otimes \mathbb{I}_2 + \sum\limits_{i=x,y,z} T_i \sigma_i \otimes \sigma_i ] $ is unsteerable if $ \max\limits_{\hat{x}} [(\vec{\mathfrak{a}}.\hat{x})^2 + 2 \| T \hat{x} \| ] \le 1 $, where $ \hat{x} $ is a unit normal vector and $ \| . \|  $ corresponds to Euclidean norm \cite{J.Bowles}. We now present below, our findings for some categories of states admitting LHS model.

\subsection{Weyl states admitting local hidden state models}
 For the Weyl states in two-qubits, the sufficient condition for unsteerability reads $ \max\limits_{\hat{x}} [ 2 \| T \hat{x} \| ] \le 1 $. Without loss of generality we take the singular values of $ T $ as $ t_1 \ge t_2 \ge t_3 $. The condition thus gives, $ t_1 \le \frac{1}{2} $. Since the conditional Rényi 2-entropy for the Weyl state is given by $  S_2 (A|B)_{\rho^{weyl}_{2}}=1-\log(1+t_{1}^2 + t_{2}^2 + t_{3}^2) $ , we have $ t_1 \le \frac{1}{2} $ implies $ S_2 (A|B)_{\rho^{weyl}_{2}} \ge 0 $. Thus, whenever Weyl states obey the sufficient criteria for unsteerability, they have a non-negative conditional Rényi 2-entropy.

\subsection{LHS models for non-Weyl states}
Consider a two-qubit non-Weyl state \cite{J.Bowles},
$\rho_2^{nl}=p|\psi_{x}\rangle\langle\psi_{x}| +(1-p)[\rho_{x}^{A}\bigotimes \frac{\mathbb{I}}{2}]$, where $|\psi_{x}\rangle = \cos x|00\rangle +\sin x|11\rangle $, $ \rho_{x}^{A} = Tr_B(|\psi_{x}\rangle \langle \psi_{x} | ) $; $0<x\leq\frac{\pi}{4}$, $0\leq p\leq 1$. Sufficient condition for unsteerability of the state is given by,
\begin{equation} \label{unnwl}
	[\cos(2x)]^2 \geq \frac{2p-1}{(2-p)p^3}.
\end{equation}
Here, we observe that $ S_2 (A|B)_{\rho^{nl}_{2}} < 0 $ for some parameter in the above range, one instance being $ p= \frac{1}{\sqrt{2}} , x = \frac{1}{2} \arccos [\sqrt{\frac{1}{7} (12-4 \sqrt{2})) }]$.

However, on mixing white noise with the state we arrive at a different observation, which we note below.

A state $\rho_2^{'}$ admits LHS model if there exists a unit trace operator $O_{AB}$ such that $O_{AB}$ admits LHS model and $\rho_2^{'}= r O_{AB} + (1-r) \frac{I}{d}\otimes O_{B}$ \cite{D.Cavalcanti}. Consider now $ O_{AB} = \rho_{2}^{nl} $.

$S_{2}(A|B)_{\rho_2^{'}}$ is non-negative for 
\begin{align}
	\cos4x\geq \frac{r^{2}+4r^{2}p^{2}-p^{2}-2}{2p^{2}-r^{2}-p^{2}(1-2r^{2})}.
\end{align}

and $ \rho_2^{'} $ is unsteerable for

\begin{align}
	\cos4x\geq \frac{2(2p-1)-(2-p)p^3}{(2-p)p^3}.
\end{align}

Let $ h(r,p)= \frac{2(2p-1)-(2-p)p^3}{(2-p)p^3}-\frac{r^{2}+4r^{2}p^{2}-p^{2}-2}{2p^{2}-r^{2}-p^{2}(1-2r^{2})}$ for $x\in(0,\frac{\pi}{4}]$,  $r\in[0,1]$, $p\in[0,1]$.\\
Then, $h(r,p)$ is non-negative, some instances being $r=0,$ $p\in \left[2-\sqrt{3},1\right]$; $r \in \left[0.132151,0.981434\right],$ $p \in \left(\sqrt{\frac{r^2}{1+2r^2}} ,1\right];$ $r=1,$ $p=1.$

Thus, implying that whenever the state $\rho_2^{'}$ admits an LHS model its 
conditional Rényi 2-entropy is non-negative, for the above restrictions on $ r $ and $ p $.

\section{LHS models in $ d \otimes d $ systems and conditional entropies}\label{LHSqudit}

\subsection{Isotropic states}
For the isotropic state in $ d \otimes d $ dimensions, given by $ \rho_d^{iso}= \eta |\psi_d \rangle \langle \psi_d |+ (1-\eta) \frac{\mathbb{I}}{d} \otimes \frac{\mathbb{I}}{d} $, where $ |\psi_d \rangle = \frac{1}{\sqrt{d}} \sum\limits_{i=0}^{d-1} | ii \rangle $, the conditional Rényi 2-entropy is non-negative for $ \eta \in (\frac{1}{d+1},\frac{1}{\sqrt{d+1}}] $.

The isotropic state admits an LHS model for projective measurements when $\eta \le \frac{H_{d}-1}{d-1}$ \cite{H.Wiseman} where $H_{d}=1+1/2 + 1/3+...+1/d$. Since, $ H_{d} \approxeq \ln(1+2d) $, after simplifying the aforementioned upper bound we obtain $ \eta \in \left(\frac{1}{d+1},\frac{\ln(1+2d)-1}{d-1}\right] $ where $\ln$ is logarithm taken in base $e$.

In the above intervals, the state is also entangled. Now if we consider the function $ f(x)= \frac{1}{\sqrt{x+1}}-\frac{\ln(1+2x)-1}{x-1} $, then $ f $ is positive for $ x \ge 3 $. Further in the case $ d=2 $, since $ \frac{1}{\sqrt{3}} > \frac{1}{2}  $, whenever the isotropic states admit an LHS model its conditional Rényi 2-entropy is non-negative. Thus, in this case the isotropic states admitting LHS model forms a subset of the states having non-negative conditional entropy.

\subsection{Werner States}
If $d_{A}=d_{B}=d$ then the state in eq.(\ref{blochqudit}) denoted by $ \rho_{d}^{g} $ has its CR2E given by, 
\begin{align}
	S_{2}(A|B)_{\rho_{d}^{g}}=\log\left[\frac{d+2||\vec{b}||^{2}}{1+\frac{2||\vec{a}||^{2}+2||\vec{b}||^{2}}{d}+\frac{4||T||^{2}}{d^2}}\right].
\end{align}
We have $S_{2}(A|B)_{\rho_{d}^{g}}$ is non-negative iff $||T||\leq \frac{1}{2}\left[\sqrt{d^{2}(d-1)-2d||\vec{a}||^{2}+2d(d-1)||\vec{b}||^{2}}\right]$.
Therefore, for the Werner state $\rho_{d}^{wer}=\frac{d-1+\eta}{d-1}\frac{I}{d^2}-\frac{\eta}{d-1}\frac{V}{d}$ where $V$ is flip operator and given by $V|ij\rangle = |ji\rangle $, CR2E is given by 
\begin{align}
	S_{2}(A|B)_{\rho_{d}^{wer}}=\log\left(\frac{d^3}{d^{2}+4||T||^2}\right).
\end{align}
Here, $||T||^{2}=\frac{1}{4}\left[\frac{\eta^{2} d^{2}(d+1)}{d-1}\right]$.\\

CR2E is non-negative for the Werner state when,
\begin{align}
	||T||\leq \frac{d\sqrt{d-1}}{2}.
\end{align} 
and it admits LHS model, when $\eta \le 1-\frac{1}{d}$ \cite{H.Wiseman}. After solving above inequality we obtained
\begin{align}
	||T||\leq \frac{\sqrt{d^{2}-1}}{2}.
\end{align}
Define a function $g(d)=\frac{d\sqrt{d-1}}{2}-\frac{\sqrt{d^{2}-1}}{2}$ for $d\geq2$.
Then $g(d)\geq 0$ for $d\geq 2$.
Thus, if the Werner state has LHS model then its CR2E is non-negative 
$ \forall d \geq 2 $.

Consider now, $\rho_d^{'}= r O_{AB} + (1-r) \frac{\mathbb{I}}{d}\otimes O_{B}$, with $ O_{AB} = \rho_{d}^{wer} $.

So,  $\rho_d^{'}$ admits LHS model for 
\begin{align}
	\eta \le 1-\frac{1}{d},
\end{align}
and 
CR2E of $\rho_d^{'}$ is non-negative for 
\begin{align}
	\eta \leq \frac{d-1}{r\sqrt{d+1}}.
\end{align}
Let $F(r,d)=\frac{d-1}{r\sqrt{d+1}}-(1-\frac{1}{d})$; $0\leq r\leq 1$; $d\geq 2$.\\
Then $F(r,d)$ is non-negative for $0 < r\leq 1$; $d\geq 2$. Thus, if the state $\rho_d^{'}$ has LHS model then its CR2E is non-negative.

\section{Conclusions}
The work here studies the relation between conditional entropies of quantum systems and their steerability. We observe that in the two-qubit scenario, states with maximally mixed marginals violate the CJWR steering inequality (a particular three-settings steering inequality) iff they have negative conditional Rényi 2-entropy. For the general two-qubit state, an upper bound to its conditional Rényi 2-entropy is given which guarantees violation of the CJWR inequality. Using a pertinent sufficiency criteria, we show that whenever Weyl states in two-qubits are unsteerable (irrespective of any particular inequality), they have a non-negative conditional Rényi 2-entropy. We extend our study to include LHS models in two-qudit systems, particularly for the isotropic and Werner states and find that whenever they admit LHS models, they possess a non-negative conditional Rényi 2-entropy. A noisy variant of the two-qudit Werner state exhibits the same trait. 

We observe that the relation between quantum conditional entropies and Bell-type inequalities is intricate where the present work suggests that no direct relation exists for a generic quantum state. However, we find that states with maximally mixed marginals do have a special role to play here pertaining to the establishment of a direct relationship. Thus, a significant area of future research could be to explore the relation between local hidden state models and conditional entropy for states with maximally mixed marginals in arbitrary $ d \otimes d $ dimensions apart from the states already discussed here. Another direction that warrants attention is to explore the relation for multipartite systems. 
\section*{Acknowledgement}

NG acknowledges support from the project grant received from SERB under the MATRICS scheme, vide file number MTR/2022/000101.


\begin{thebibliography}{1}
	
	\bibitem{P.Zoller}Zoller, P.,\textit{et.al.}, \textit{Quantum information processing and communication}, \href{https://doi.org/10.1140/epjd/e2005-00251-1}{The European Physical Journal D, \textbf{36}, 203(2005).}
	
	\bibitem{R1.Horodecki}Horodecki, R., Horodecki, P., Horodecki, M. and Horodecki, K., \textit{Quantum entanglement}, \href{https://doi.org/10.1103/RevModPhys.81.865}{Reviews of Modern Physics, \textbf{81}, 865(2009).}
	
	\bibitem{J.Bell}Bell, J.S., \textit{On the Einstein Podolsky Rosen paradox}, \href{https://doi.org/10.1103/PhysicsPhysiqueFizika.1.195}{Physics Physique Fizika, \textbf{1}, 195(1964).}
	
	
	\bibitem{V.Scarani} Brunner,B.,Cavalcanti,D., Pironio,S., Scarani,V. and  Wehner,S., \textit{Bell nonlocality}, \href{https://doi.org/10.1103/RevModPhys.86.419}{Reviews of Modern Physics 86,419 (2014)};
	Scarani, V., \textit{Bell nonlocality}, Oxford University Press, 2019.
	
	\bibitem{A.Acin}Acín, A., Brunner, N., Gisin, N., Massar, S., Pironio, S. and Scarani, V., \textit{Device-independent security of quantum cryptography against collective attacks}, 
	\href{https://doi.org/10.1103/PhysRevLett.98.230501}{Physical Review Letters, \textbf{98}, 230501(2007).}

   \bibitem{R.Uola}Uola, R., Costa, A.C., Nguyen, H.C. and Gühne, \textit{Quantum steering}, \href{https://doi.org/10.1103/RevModPhys.92.015001}{Reviews of Modern Physics, \textbf{92}, 015001(2020).}

   \bibitem{E.Schrodinger1}Schrödinger, E., \textit{Discussion of probability relations between separated systems},\\ \href{https://doi.org/10.1017/S0305004100013554}{In Mathematical Proceedings of the Cambridge Philosophical Society, Cambridge University Press, \textbf{31}, 555(1935).}
   
   \bibitem{E.Schrodinger2}Schrödinger, E., \textit{Probability relations between separated systems},\\ \href{https://doi.org/10.1017/S0305004100019137}{In Mathematical Proceedings of the Cambridge Philosophical Society, Cambridge University Press, \textbf{32}, 446­(1936).}

   \bibitem{H.Wiseman}Wiseman, H.M., Jones, S.J. and Doherty, A.C., \textit{Steering, entanglement, nonlocality, and the Einstein-Podolsky-Rosen paradox}, \href{https://doi.org/10.1103/PhysRevLett.98.140402}{Physical Review Letters, \textbf{98}, 140402(2007).}
   
   \bibitem{M.Pawlowski}Pawłowski, M. and Brunner, N., \textit{Semi-device-independent security of one-way quantum key distribution}, \href{https://doi.org/10.1103/PhysRevA.84.010302}{Physical Review A, \textbf{84}, 010302R (2011).}


   \bibitem{A.Wehrl}Wehrl, A., \textit{General properties of entropy}, \href{https://doi.org/10.1103/RevModPhys.50.221}{Reviews of Modern Physics, \textbf{50}, 221(1978).}

   \bibitem{C.E.Shannon}Shannon, C.E., \textit{A mathematical theory of communication}, \href{https://doi.org/10.1002/j.1538-7305.1948.tb01338.x}{The Bell System Technical journal, \textbf{27}, 379(1948).}

   \bibitem{J.Neumann}Neumann,von J.,\textit{Mathematical foundations of quantum mechanics}, Princeton University Press, 2018.

   \bibitem{AR}Rényi, A., \textit{On measures of entropy and information}, In Proceedings of the fourth Berkeley symposium on mathematical statistics and probability, \textbf{1}, 547(1961).
  

   \bibitem{TS}Tsallis, C., \textit{Possible generalization of Boltzmann-Gibbs statistics}, \href{https://doi.org/10.1007/BF01016429}{Journal of Statistical Physics, \textbf{52}, 479(1988).}

   \bibitem{M.Muller}Müller-Lennert, M., Dupuis, F., Szehr, O., Fehr, S. and Tomamichel, M., \textit{On quantum Rényi entropies: A new generalization and some properties}, \href{https://doi.org/10.1063/1.4838856}{Journal of Mathematical Physics, \textbf{54}, 122203(2013).}


   \bibitem{N.Cerf}Cerf, N.J. and Adami, C., \textit{Entropic Bell inequalities}, \href{https://doi.org/10.1103/PhysRevA.55.3371}{Physical Review A, \textbf{55}, 3371(1997).}

   \bibitem{J.Schneeloch}Schneeloch, J., Broadbent, C.J., Walborn, S.P., Cavalcanti, E.G. and Howell, J.C., \textit{Einstein-Podolsky-Rosen steering inequalities from entropic uncertainty relations}, \href{https://doi.org/10.1103/PhysRevA.87.062103}{Physical Review A, \textbf{87}, 062103(2013).}
   
   \bibitem{AC}Costa, A.C., Uola, R. and Gühne, O., \textit{Entropic steering criteria: applications to bipartite and tripartite systems}, \href{https://doi.org/10.3390/e20100763}{Entropy, \textbf{20}, 763(2018).}
 
   \bibitem{S.Wehner}Wehner, S. and Winter, A., \textit{Entropic uncertainty relations—a survey}, \href{https://doi.org/10.1088/1367-2630/12/2/025009}{New Journal of Physics, \textbf{12}, 025009(2010).}; 
   Coles,P.J., Berta,M., Tomamichel,M. and Wehner,S., \textit{Entropic uncertainty relations and their applications}, \href{https://doi.org/10.1103/RevModPhys.89.015002}{Reviews of Modern Physics, \textbf{89}, 015002(2017).}

   \bibitem{N.Gisin}Gisin, N., Ribordy, G., Tittel, W. and Zbinden, H., \textit{Quantum cryptography}, \href{https://doi.org/10.1103/RevModPhys.74.145}{Reviews of Modern Physics, \textbf{74}, 145(2002).}

   \bibitem{N.J.Cerf}Cerf, N.J. and Adami, C., \textit{Negative entropy and information in quantum mechanics}, \href{https://doi.org/10.1103/PhysRevLett.79.5194}{Physical Review Letters, \textbf{79}, 5194(1997).}
   
   \bibitem{M.Horodecki}Horodecki, M., Oppenheim, J. and Winter, A., \textit{Quantum state merging and negative information}, \href{https://doi.org/10.1007/s00220-006-0118-x}{Communications in Mathematical Physics, \textbf{269}, 107(2007).}
   
   \bibitem{SLP}Slepian, D. and Wolf, J., \textit{Noiseless coding of correlated information sources}, \href{https://doi.org/10.1109/TIT.1973.1055037}{IEEE Transactions on Information Theory, \textbf{19}, 471(1973).}

   \bibitem{M1.Vempati}Vempati, M., Ganguly, N., Chakarabarty, I. and Pati, A.K., \textit{Witnessing negative conditional entropy}, \href{https://doi.org/10.1103/PhysRevA.104.012417}{Physical Review A, \textbf{104}, 012417(2021).}

   \bibitem{M2.Vempati}Vempati, M., Shah, S., Ganguly, N.  and Chakrabarty, I., \textit{A-unital Operations and Quantum Conditional Entropy}, \href{https://doi.org/10.22331/q-2022-02-02-641}{Quantum, \textbf{6}, 641(2022).}
   
   \bibitem{MW}Brandsen, S., Geng, I.J., Wilde, M.M. and Gour, G., \textit{Quantum conditional entropy from information-theoretic principles}, \href{https://arxiv.org/abs/2110.15330}{arXiv preprint,arXiv:2110.15330.}

   \bibitem{Gourinev} Gour, G., Wilde, M.M.,Brandsen, S. and Geng, I.J., \textit{Inevitability of knowing less than nothing},\href{https://doi.org/10.48550/arXiv.2208.14424}{arXiv preprint,arXiv:2208.14424.}    
   
   \bibitem{J.F.Clauser}Clauser, J.F., Horne, M.A., Shimony, A. and Holt, R.A., \textit{Proposed experiment to test local hidden-variable theories}, \href{https://doi.org/10.1103/PhysRevLett.23.880}{Physical Review Letters, \textbf{23}, 880(1969).}

   \bibitem{N.Friis}Friis, N., Bulusu, S. and Bertlmann, R.A., \textit{Geometry of two-qubit states with negative conditional entropy}, \href{https://doi.org/10.1088/1751-8121/aa5dfd}{Journal of Physics A: Mathematical and Theoretical, \textbf{50}, 125301(2017).}

   \bibitem{R.Horodecki} Horodecki, R., Horodecki, P. and Horodecki, M., \textit{Quantum  $\alpha$-entropy inequalities: independent condition for local realism?}, \href{https://doi.org/10.1016/0375-9601(95)00930-2}{Physics Letters A, \textbf{210}, 377(1996).}

   \bibitem{E.Cavalcanti}Cavalcanti, E.G., Jones, S.J., Wiseman, H.M. and Reid, M.D., \textit{Experimental criteria for steering and the Einstein-Podolsky-Rosen paradox}, \href{https://doi.org/10.1103/PhysRevA.80.032112}{ Physical Review A, \textbf{ 80}, 032112(2009).}

   \bibitem{J.Bowles}Bowles, J., Hirsch, F., Quintino, M.T. and Brunner, N., \textit{Sufficient criterion for guaranteeing that a two-qubit state is unsteerable}, \href{https://doi.org/10.1103/PhysRevA.93.022121}{Physical Review A, \textbf{93}, 022121(2016).}
   
   \bibitem{R.Werner}Werner, R.F., \textit{Quantum states with Einstein-Podolsky-Rosen correlations admitting a hidden-variable model}, \href{https://doi.org/10.1103/PhysRevA.40.4277}{Physical Review A, \textbf{40}, 4277(1989).}
   
   \bibitem{Vollbrecht}Vollbrecht, K.G.H. and Wolf, M.M., \textit{Conditional entropies and their relation to entanglement criteria}, \href{https://doi.org/10.1063/1.1498490}{Journal of Mathematical Physics, \textbf{43}, 4299(2002).}
   
   \bibitem{HoroBell} Horodecki,R., Horodecki,P. and Horodecki,M., \textit{Violating Bell inequality by mixed spin 1/2 states:necessary and sufficient condition},\href{https://doi.org/10.1016/0375-9601(95)00214-N}{Physics Letters A 200, 340(1995)}.

   \bibitem{Costa}Costa, A.C.S. and Angelo, R.M., \textit{Quantification of Einstein-Podolsky-Rosen steering for two-qubit states}, \href{https://doi.org/10.1103/PhysRevA.93.020103}{Physical Review A, \textbf{93}, 020103(2016).}

   \bibitem{S.Bhattacharya}Bhattacharya, S.S., Mukherjee, A., Roy, A., Paul, B., Mukherjee, K., Chakrabarty, I., Jebaratnam, C. and Ganguly, N., \textit{Absolute non-violation of a three-setting steering inequality by two-qubit states}, \href{https://doi.org/10.1007/s11128-017-1734-4}{Quantum Information Processing, \textbf{17}, 3(2018).}

   \bibitem{D.Cavalcanti}Cavalcanti, D., Guerini, L., Rabelo, R. and Skrzypczyk, P., \textit{General method for constructing local hidden variable models for entangled quantum states}, \href{https://doi.org/10.1103/PhysRevLett.117.190401}{Physical Review Letters, \textbf{117}, 190401(2016).}

\end{thebibliography}
\end{document}